\documentclass[useAMS,usenatbib]{mn2e}

\title[Shear and star formation in spiral galaxies]{The connection between shear and star formation in spiral galaxies}
\author[Marc S. Seigar]{Marc S. Seigar\thanks{E-mail:
mseigar@uci.edu}\\Department of Physics \& Astronomy, University of California Irvine, 4129 Frederick Reines Hall, Irvine, CA 92697-4575, USA}
\begin{document}

\date{Accepted 2005 April 24.  Received 2005 April 22; in original form 2005 March 30.}

\pagerange{\pageref{firstpage}--\pageref{lastpage}} \pubyear{2005}

\maketitle

\label{firstpage}

\begin{abstract}
We present a sample of 33 galaxies for which we have calculated (i) the 
average rate of shear from publish rotation curves, (ii) the far--infrared 
luminosity from IRAS fluxes and (iii) The K--band luminosity from 2MASS. We 
show that a correlation exists between the shear rate and the ratio of the 
far--infrared to K--band luminosity. This ratio is essentially a measure of 
the star formation rate per unit mass, or the specific star formation rate. 
From this correlation we show that a critical shear rate exists, above
which star formation would turn off in the disks of spiral galaxies. Using
the correlation between shear rate and spiral arm pitch angle, this shear
rate corresponds to the lowest pitch angles typically measured in 
near-infrared images of spiral galaxies.
\end{abstract}

\begin{keywords}
galaxies: fundamental parameters -- galaxies: spiral -- infrared: galaxies
\end{keywords}

\section{Introduction}

Jeans (1929) took up the question of self--gravitating gas and found
that under certain conditions it could be unstable enough to collapse
under its own self--gravity. In order to show this he considered an
adiabatic gas. A similar calculation can be applied to stellar systems.

The idea of instabilities is essentially a condensation of material,
so in order for material to condense it is necessary to find out if
gravity will cause collapse before velocity dispersion 
causes expansion. A characteristic time can be calculated for each
process and compared to see which process is dominant. It turns out
that on small scales velocity dispersion is dominant, and on large
scales gravity is dominant. Jeans (1929) found a critical length
above which gravitational instability becomes dominant, now known
as the Jeans length, $L_{Jeans}$.

The situation in the disks of galaxies is different from the 
problem formulated by Jeans, due to the flatness of the system
(instead of a spheroid assumed in the Jeans analysis) and more
importantly, differential rotation. Velocities due to differential
rotation are approximately proportional to $\Delta R$ and might
even prevent the collapse from taking place, even at distances
greater than the Jeans length. Differential rotation inhibits the
gravitational collapse on large scales. The question is what 
happens in between? Toomre (1964) attempted to answer this 
question. He investigated the balance between 
differential rotation and self--gravitation. Differential
rotation manifests itself physically from the fact that a 
contracting region conserves angular momentum. This spins up and
causes a centrifugal force that might inhibit further collapse.
Toomre (1964) found that differential rotation implies that 
disks are stabilised at lengths greater than a critical length, which
we will call the rotation length, $L_{rot}$. It turns out that
$L_{rot}$ is inversely proportional to the average angular velocity
with respect to a fixed system. The higher the average angular
velocity, the greater the rate of shear, $A/\omega$. In other
words as the rate of shear increases, the value of $L_{rot}$ decreases, 
until it reaches the limiting case where $L_{rot}=L_{Jeans}$, and the
entire disk is stable against gravitational collapse, i.e no clouds
will collapse and therefore no star formation can occur.

In reality most disks have a region where they are unstable to gravitational
instabilities and this region has a length scale between the Jeans length
and the rotation length, i.e. $L_{Jeans} < L < L_{rot}$. However, the 
faster a disk rotates, the stronger its differential rotation. The stronger
differential rotation becomes, the more inhibited star formation becomes,
and thus we expect a correlation between star formation rate and shear rate.

Toomre (1964) derived a value of the velocity dispersion 
for the limiting regime where $L_{rot}=L_{Jeans}$, i.e. a critical velocity
dispersion. 
The stability of disks can then be quoted as the ratio of of the actual 
velocity dispersion to the critical velocity dispersion, $Q$.
Thus if $Q > 1$ then the velocity dispersion is high enough to prevent
gravitational collapse, and if $Q < 1$ then gravitational collapse 
occurs. This condition is known as the {\em Toomre stability criterion}
and the parameter $Q$ is known as {\em Toomre's parameter}.

Since differential rotation acts to shear features in the disk, the
shear rate is a viable quantity for measuring the differential 
rotation in the disks of spiral galaxies. Shear rate has been
measured by several authors (e.g. Block et al. 1999; Seigar, Block \& Puerari
2004; Seigar et al. 2005) and can be measured directly from a galaxy rotation
curve. The sequence of
low shear rates to high shear rates also follows the sequence
from late-type spirals to early-type spirals (Seigar et al. 2005).
It has been shown that as one progresses along the Hubble 
sequence, the specific star formation rate (measured as the 
H$\alpha$ equivalent width) in galaxies increases
(e.g. James et al. 2004). If one takes this along
with the correlation between pitch angle and shear rate in
Seigar et al. (2005), then the existence of a correlation
between shear rate and specific star formation rate is implied,
assuming that the transition from tightly wound spiral structure
to losely wound spiral structure follows the Hubble sequence
from early- to late-type spiral galaxies.

This letter is arranged as follows. Section 2 describes the data we
have used. Section 3 describes how shear rates were calculated using
rotation curves. Section 4 describes a method for determining the
star formation rate per unit surface area. Finally, section 5 discusses
the results.

\section{Data}

In order to calculate shear rates, rotation curves are required. Burstein
\& Rubin (1985) presented data for 60 galaxies for which they had observed 
rotation curves. The rotation curve data was available in a series of papers
from the early 1980s (Rubin, Ford \& Thonnard 1980; Rubin et al. 1982, 1985).
Of this sample of 60 galaxies, 40 were detected by IRAS at 60$\mu$m and
100$\mu$m, and by 2MASS in the $K_s$-band. 
6 of these galaxies were rejected from our sample, as they were
classified as HII or starburst and may therefore have external physical 
processes affecting their star formation rates. A further  object did
not have a rotation curve that covered a sufficient radial range and was also
not included in the sample. The remaining 33 galaxies are presented here.

\begin{table*}
\caption[]{Data for the 33 galaxies. Column 1 lists the name of the galaxy; Column 2 lists the Hubble type from de Vaucouleurs et al. (1991; hereafter RC3); Column 3 lists the shear rates as calculated from the rotation curves; Column 4 lists the IRAS 60$\mu$m flux; Column 5 lists the IRAS 100$\mu$m flux; Column 6 lists the distance to the objects in Mpc; Column 7 lists the diameters out to the $\mu_R=25$ mag arcsec$^{-2}$ isophote ($D_{25}$) from RC3; Column 8 lists the far--infrared luminosity in solar units calculated using equation 2; Column 9 lists the apparent K$_s$-band total magnitude from 2MASS; Column 10 lists the K-band luminosity calculated using equation 3 and column 11 lists the inferred star formation rate.}
\centering
\begin{tabular}{llllllllllll}
\hline\hline
Galaxy		& Hubble        & Shear		& $S_{60}$& $S_{100}$ 	& $D$	& $D_{25}$ & $L_{FIR}$			& $K_T$		& $L_{K}$	& SFR	\\
name		& Type          & rate		& (Jy)	& (Jy)		& (Mpc)	& (kpc)    & ($\times 10^9$L$_\odot$)   &	        & ($\times 10^{10}$L$_{\odot}$)& ($M_{\odot}/year$)\\
\hline
NGC 753		& SABbc         & 0.38$\pm$0.03	& 3.36	& 11.40 & 65.4	& 47.79$\pm$2.25 & 32.17$\pm$2.25& 9.37$\pm$0.02 & 17.88$\pm$0.33	& 8.30$\pm$1.71\\
NGC 801		& Sc            & 0.53$\pm$0.03	& 1.45	& 5.06	& 76.9	& 70.73$\pm$5.05 & 19.47$\pm$1.27& 9.51$\pm$0.03 & 21.82$\pm$0.59	& 6.20$\pm$1.19\\
NGC 1024 	& SAab          & 0.65$\pm$0.02 & 0.57	& 2.62	& 47.1	& 53.30$\pm$2.54 & 3.40$\pm$0.49 & 8.74$\pm$0.02 & 16.49$\pm$0.30	& 2.31$\pm$0.43\\
NGC 1035	& SAc?          & 0.37$\pm$0.03	& 3.57	& 11.12 & 16.6	& 10.81$\pm$0.77 & 2.09$\pm$0.13 & 9.13$\pm$0.01 & 1.42$\pm$0.01	& 0.68$\pm$0.14\\
NGC 1085	& SAbc          & 0.52$\pm$0.04	& 0.88	& 3.16	& 90.5	& 77.69$\pm$7.50 & 16.71$\pm$1.25& 9.80$\pm$0.03 & 23.40$\pm$0.64	& 6.90$\pm$1.40\\
NGC 1357	& SAab          & 0.56$\pm$0.03	& 0.93	& 4.67	& 26.8	& 21.97$\pm$1.57 & 1.90$\pm$0.21 & 8.42$\pm$0.03 & 7.18$\pm$0.20	& 1.78$\pm$0.34\\
NGC 1417	& SABb          & 0.45$\pm$0.03	& 1.59	& 5.82	& 54.2	& 42.43$\pm$3.03 & 10.95$\pm$0.60& 9.14$\pm$0.03 & 15.28$\pm$0.42	& 5.81$\pm$1.15\\
NGC 1421	& SABbc         & 0.31$\pm$0.04	& 8.48	& 21.32 & 27.8	& 28.69$\pm$1.35 & 12.55$\pm$0.69& 8.40$\pm$0.02 & 7.91$\pm$0.14	& 4.34$\pm$1.05\\
NGC 1620	& SABbc         & 0.44$\pm$0.03	& 1.31	& 5.31	& 46.8	& 39.26$\pm$2.80 & 7.15$\pm$0.43 & 8.92$\pm$0.03 & 13.91$\pm$0.38	& 5.46$\pm$1.08\\
NGC 2590	& SAbc          & 0.44$\pm$0.03	& 2.01	& 6.03	& 66.6	& 43.38$\pm$3.08 & 18.68$\pm$0.93& 9.38$\pm$0.03 & 18.43$\pm$0.50	& 7.24$\pm$1.44\\
NGC 2608	& SBb           & 0.45$\pm$0.04	& 2.25	& 5.77	& 28.5	& 18.99$\pm$0.90 & 3.51$\pm$0.23 & 9.33$\pm$0.03 & 3.51$\pm$0.10	& 1.33$\pm$0.28\\
NGC 2639	& SAa?          & 0.54$\pm$0.03	& 1.99	& 7.06	& 44.5	& 23.56$\pm$2.87 & 9.04$\pm$0.36 & 8.40$\pm$0.03 & 20.25$\pm$0.55	& 5.51$\pm$1.07\\
NGC 2742	& SAc           & 0.40$\pm$0.03	& 3.08	& 10.49 & 17.2  & 15.11$\pm$1.08 & 2.04$\pm$0.10 & 8.81$\pm$0.01 & 2.06$\pm$0.02	& 0.91$\pm$0.18\\
NGC 2775	& SAab          & 0.63$\pm$0.03	& 1.80	& 9.46	& 18.1	& 22.46$\pm$1.06 & 1.74$\pm$0.12 & 7.04$\pm$0.02 & 11.60$\pm$0.21	& 1.90$\pm$0.36\\
NGC 2815	& SBb           & 0.60$\pm$0.03	& 1.04	& 4.79	& 33.9	& 34.19$\pm$1.03 & 3.21$\pm$0.27 & 8.25$\pm$0.03 & 13.44$\pm$0.37	& 2.68$\pm$0.52\\
NGC 2844        & SAa           & 0.63$\pm$0.03 & 0.41  & 1.91  & 19.8  & 8.92$\pm$0.64  & 0.44$\pm$0.06 & 9.89$\pm$0.03 & 1.01$\pm$0.03       & 0.16$\pm$0.04\\
NGC 2998	& SABc          & 0.41$\pm$0.03	& 1.58	& 4.63	& 63.8	& 53.53$\pm$2.52 & 13.28$\pm$0.93& 9.93$\pm$0.04 & 10.23$\pm$0.37	& 4.39$\pm$0.89\\
NGC 3223	& SAbc          & 0.61$\pm$0.03	& 3.79	& 14.76 & 38.6	& 45.75$\pm$2.16 & 13.71$\pm$1.51& 7.58$\pm$0.02 & 32.22$\pm$0.59	& 6.06$\pm$1.16\\
NGC 3281	& SABa          & 0.45$\pm$0.02	& 6.86	& 7.51	& 42.7	& 41.13$\pm$2.94 & 17.21$\pm$1.20& 8.31$\pm$0.03 & 20.19$\pm$0.55	& 7.68$\pm$1.44\\
NGC 3495	& Sd            & 0.42$\pm$0.02	& 1.82	& 7.28	& 15.2	& 21.66$\pm$1.02 & 1.03$\pm$0.08 & 8.93$\pm$0.02 & 1.43$\pm$0.03	& 0.60$\pm$0.11\\
NGC 3672	& SAc           & 0.43$\pm$0.02	& 7.33	& 20.80 & 24.8	& 30.08$\pm$1.42 & 9.18$\pm$0.78 & 8.27$\pm$0.01 & 7.08$\pm$0.06	& 2.86$\pm$0.53\\
NGC 4378	& SAa           & 0.69$\pm$0.03	& 0.36	& 1.45	& 34.1	& 28.61$\pm$1.35 & 1.04$\pm$0.18 & 8.51$\pm$0.02 & 10.71$\pm$0.20	& 1.00$\pm$0.18\\
NGC 4682	& SABcd         & 0.51$\pm$0.03	& 0.69	& 2.38	& 31.1	& 23.25$\pm$1.66 & 1.51$\pm$0.14 & 9.60$\pm$0.02 & 3.25$\pm$0.06	& 1.00$\pm$0.20\\
NGC 6314	& SAa           & 0.60$\pm$0.03	& 0.51	& 2.85	& 88.4	& 37.16$\pm$3.60 & 12.18$\pm$1.10& 9.81$\pm$0.03 & 22.12$\pm$0.60	& 4.43$\pm$0.84\\
NGC 7083	& SABc          & 0.43$\pm$0.03	& 5.02	& 17.19 & 41.5	& 49.96$\pm$2.36 & 19.41$\pm$0.87& 8.42$\pm$0.03 & 17.17$\pm$0.47	& 6.95$\pm$1.39\\
NGC 7171	& SBb           & 0.47$\pm$0.03	& 0.89	& 3.33	& 36.3	& 27.77$\pm$1.31 & 2.77$\pm$0.18 & 9.31$\pm$0.04 & 5.78$\pm$0.21	& 2.06$\pm$0.41\\
NGC 7217	& SAab          & 0.66$\pm$0.02	& 4.96	& 18.45 & 12.7	& 14.37$\pm$0.68 & 1.89$\pm$0.17 & 6.83$\pm$0.01 & 6.93$\pm$0.06	& 0.89$\pm$0.16\\
IC 467		& SABc          & 0.50$\pm$0.04	& 0.89	& 3.16	& 27.2	& 25.60$\pm$1.21 & 1.52$\pm$0.10 & 10.04$\pm$0.05& 1.66$\pm$0.07	& 0.53$\pm$0.11\\
IC 724		& Sa            & 0.69$\pm$0.03	& 0.34	& 1.27	& 79.6	& 54.28$\pm$5.23 & 5.09$\pm$0.81 & 9.39$\pm$0.04 & 26.10$\pm$0.94	& 2.40$\pm$0.44\\
UGC 3691	& SAcd          & 0.33$\pm$0.03	& 1.36	& 3.79	& 29.4	& 18.71$\pm$1.80 & 2.36$\pm$0.21 & 10.31$\pm$0.07& 1.51$\pm$0.09	& 0.80$\pm$0.16\\
UGC 10205	& Sa            & 0.48$\pm$0.03	& 0.39	& 1.54	& 87.4	& 36.74$\pm$4.50 & 7.32$\pm$1.02 & 9.89$\pm$0.03 & 19.98$\pm$0.54	& 6.88$\pm$1.35\\
UGC 11810	& SABbc         & 0.42$\pm$0.02	& 0.50	& 2.04	& 63.0	& 33.35$\pm$1.57 & 4.93$\pm$0.57 & 10.82$\pm$0.06& 4.38$\pm$0.24	& 1.81$\pm$0.34\\
UGC 12810	& SABbc         & 0.44$\pm$0.02	& 0.78	& 1.78	& 108.2	& 58.61$\pm$5.63 & 16.59$\pm$1.99& 10.47$\pm$0.06& 18.04$\pm$0.97	& 7.08$\pm$1.33\\
\hline
\end{tabular}
\label{landtable}
\end{table*}

\section{Calculation of shear rate}

The rotation curves presented by Rubin et al. (1980, 1982, 1985) are of 
good quality and can be used to derive shear rates. The rates of shear are
derived from their rotation curves as follows,
\begin{equation}
\label{shear}
\frac{A}{\omega}=\frac{1}{2}\left(1-\frac{R}{V}\frac{dV}{dR}\right)
\end{equation}
where $A$ is the first Oort constant, $\omega$ is the angular
velocity and $V$ is the measured line--of--sight velocity at a 
radius $R$. The value of $A/\omega$ is the shear rate.

Using equation (\ref{shear}), we have calculated mean shear rates for these
galaxies, over a radial range with the inner limit near the turnover radius,
and the outer limit at the $D_{25}$ radius. The dominant source of error on the
shear rate is the spectroscopic errors (i.e. a combination of the intrinsic
spectroscopic error and the error associated with folding the two sides of
the galaxy), which Rubin et al. (1980, 1982, 1985) claim is quite small,
typically $<10$ per cent. In order to calculate the shear rate, the mean 
value of $dV/dR$ measured in km s$^{-1}$ arcsec$^{-1}$ is calculated by
fitting a line of constant gradient to the outer part of the rotation curve
(i.e. past the radius of rotation). This results in a mean shear rate.
This is essentially the same method used by other authors to calculate
shear rate (e.g. Block et al. 1999; Seigar, Block \& Puerari 2004; Seigar 
et al. 2005). The shear rates for these galaxies are listed in Table 1.

\section{IRAS fluxes as an indicator of star formation rates}

The IRAS 60$\mu$m and 100$\mu$m fluxes have been used to calculate the 
far--infared luminosity for the galaxies in this sample for which IRAS data
was available (see Table 1). 
The far--infrared luminosity $L_{FIR}$ can be used as a quantitative
indicator of star formation rates (Spinoglio et al. 1995; Seigar \& James 
1998a). This was divided by the K-band luminosity $L_{K}$ of the galaxies to 
compensate for differences in their overall size. The $K$ band luminosity
can be used to estimate the stellar mass in galaxies (e.g. Brinchmann
\& Ellis 2000; Gavazzi et al. 2002) and so the $L_{FIR}/L_{K}$ ratio can
be interpretted as a star formation rate per unit stellar mass, i.e. a 
specific star formation rate. This is closely related to the birthrate
parameter (Gavazzi et al. 2002), i.e. the fraction of young to old
stars. The far--infared luminosity
in terms of the 60$\mu$m and 100$\mu$m flux is given by Lonsdale et al. 
(1985) as,
\begin{equation}
\label{lonsdale}
L_{FIR}=3.75 \times 10^{5}D^{2}(2.58S_{60}+S_{100})
\end{equation}
where $L_{FIR}$ is the far--infrared luminosity in solar units, $D$ is the
distance to the galaxy in Mpc, $S_{60}$ is the 60$\mu$m flux in Jy and 
$S_{100}$ is the 100$\mu$m flux in Jy. As these galaxies are nearby, for 
the calculattion of distance $D$, a simple Hubble flow is assumed and a Hubble
constant, $H_{0}=75$ km s$^{-1}$ Mpc$^{-1}$, is adopted.

Given an apparent K-band magnitude it is possible to calculate a K-band
luminosity using a relationship from Seigar \& James (1998a),
\begin{equation}
\label{seigar}
\log_{10}{L_{K}}=11.364-0.4K_{T}+\log_{10}{(1+z)}+2\log_{10}{D}
\end{equation}
where $L_{K}$ is the K-band luminosity in solar units, $K_{T}$ is the K-band
apparent magnitude and $z$ is the redshift of the galaxy. The 
$\log_{10}{(1+z)}$ term is a first order K-correction. In order to calculate
this total K-band luminosity, apparent K-band magnitudes from the 2MASS survey 
were used.

\section{Discussion}

\begin{figure}
\includegraphics{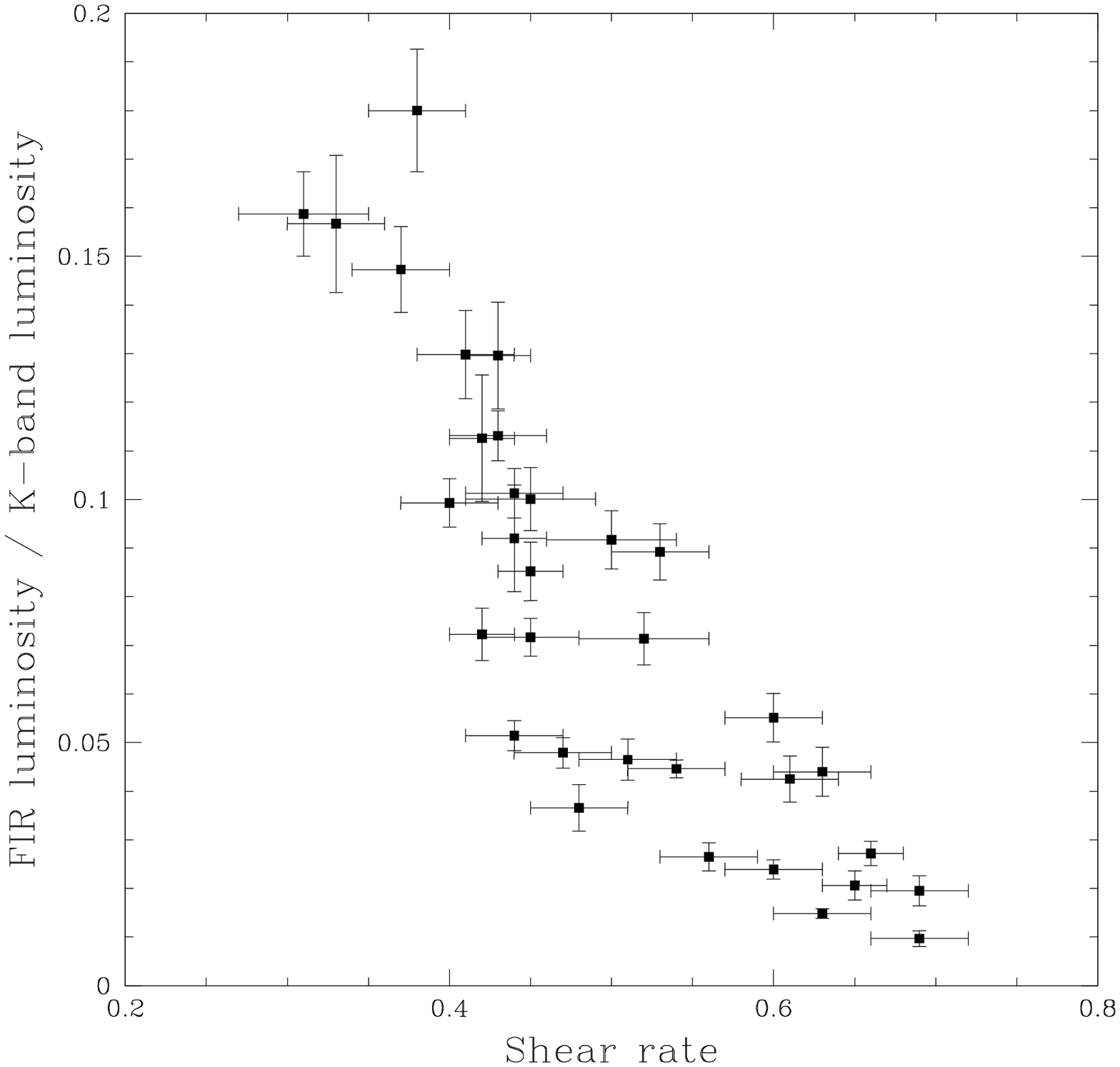}
\vspace*{8.2cm}
\caption{Shear rate, $A/\omega$, versus ratio of FIR to K-band luminosity, $L_{FIR}/L_{K}$}
\end{figure}

\begin{figure}
\includegraphics{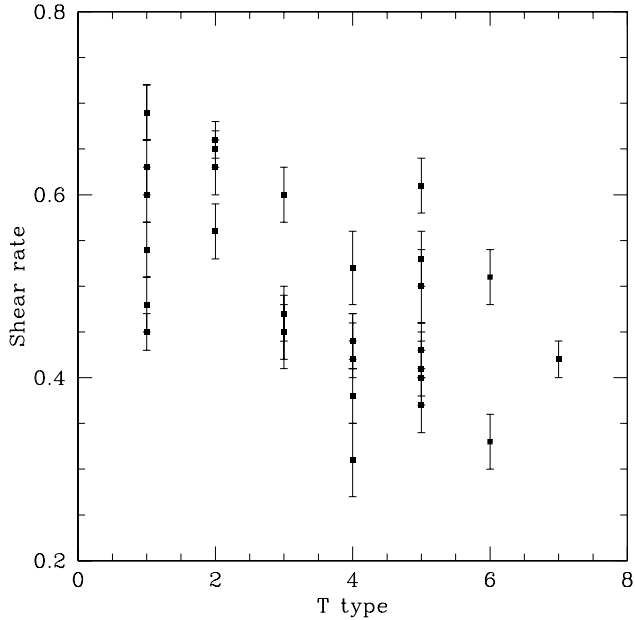}
\vspace*{8.2cm}
\caption{T type versus shear rate}
\end{figure}

\begin{figure}
\includegraphics{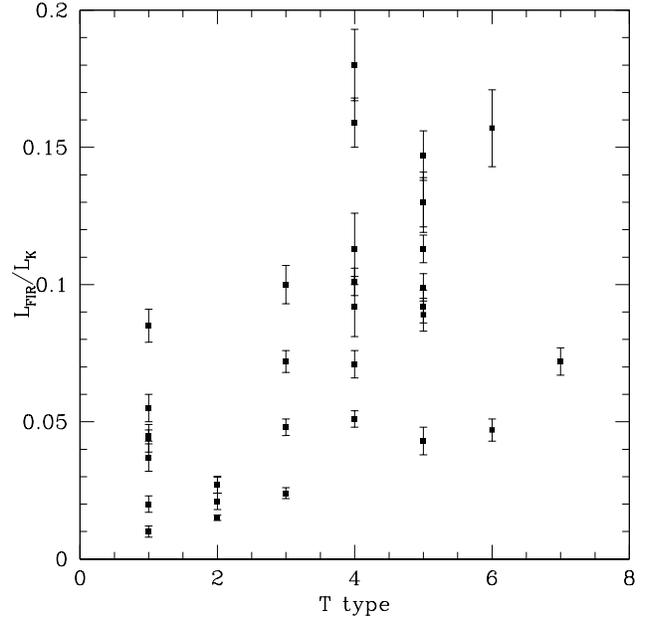}
\vspace*{8.2cm}
\caption{T type versus ratio of FIR to K-band luminosity, $L_{FIR}/L_{K}$}
\end{figure}

\begin{figure}
\includegraphics{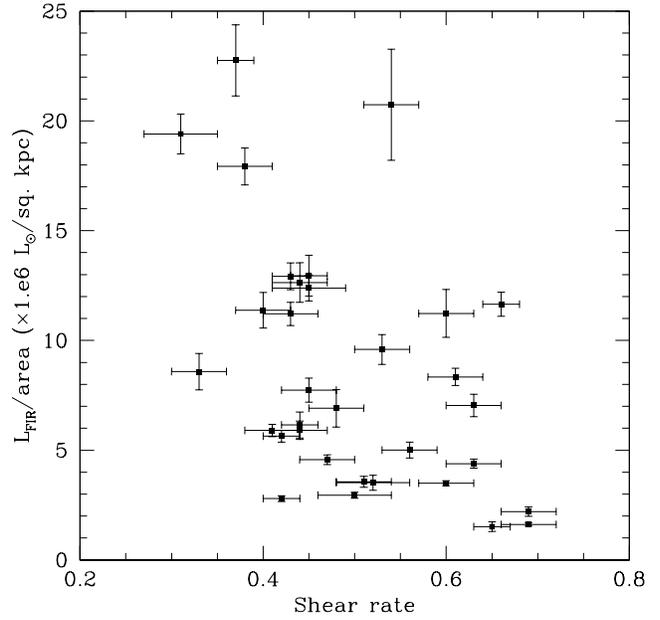}
\vspace*{8.2cm}
\caption{Shear rate, $A/\omega$, versus star formation rate per unit area in $M_{\odot}$ year$^{-1}$ kpc$^{-2}$.}
\end{figure}

\begin{figure}
\includegraphics{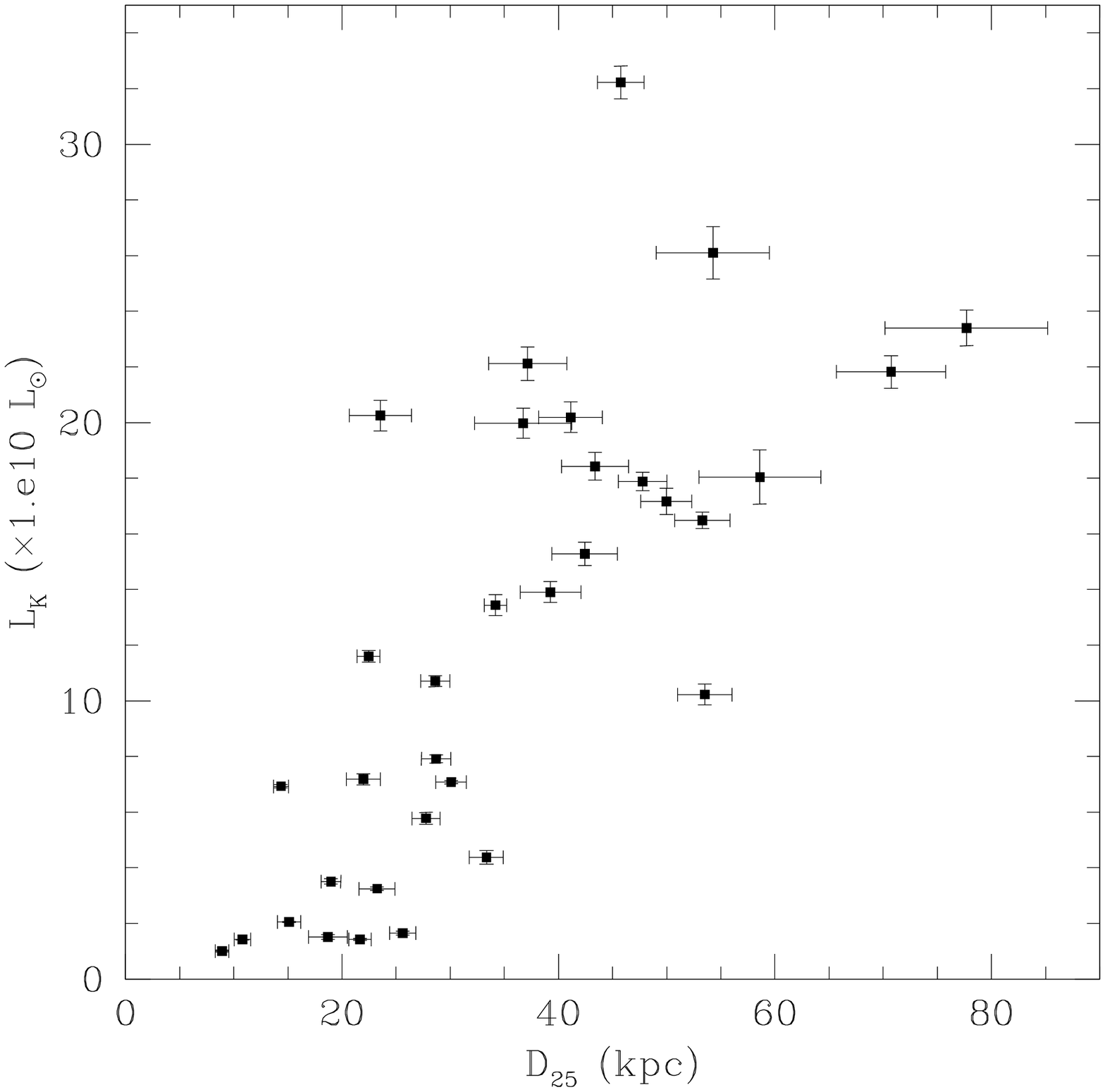}
\vspace*{8.2cm}
\caption{$D_{25}$ diameter in kpc versus 2MASS $K_s$ band luminosity.}
\end{figure}

Fig.\ 1 shows a plot of the ratio of far--infrared luminosity
to K-band luminosity versus shear rate. A good correlation
is shown (correlation coefficient = 0.71; significance = 
99.91\%). This is essentially a correlation between the
star formation rate per unit stellar mass (or the specific
star formation rate) and shear rate.
This correlation has been used to determine the relationship
between shear rate and the far--infrared luminosity as 
follows,
\begin{equation}
\label{correlation}
\frac{L_{FIR}}{L_{K}}=(0.269\pm0.020)-(0.386\pm0.040)\left(\frac{A}{\omega}\right)
\end{equation}
Where $L_{FIR}$ is the far--infrared luminosity, $L_K$ is
the K--band luminosity and $A/\omega$ is the shear rate.
The relationship between star formation rate (in 
$M_{\odot}/$year) and the far--infrared luminosity (in ergs/s)
for normal spiral galaxies is given by Buat \& Xu (1996) as
\begin{equation}
\label{kenn1}
SFR(M_{\odot}/year)=8\times 10^{-44} L_{FIR} (ergs/s)
\end{equation}
This has been used to calculate the star formation rates listed in Table 1.

Galaxies with higher shear rates have tighter spiral structure. These are the
early--type galaxies (Seigar, Block \& Puerari 2004; Seigar et al. 2005). 
Those with low shear rates have losely wound spiral structure. These are
the late--type galaxies. The transition from early--type to late--type
spiral galaxies also follows a transition from galaxies with low amounts of
gas to gas--rich galaxies (Bertin 1991; 
Block et al. 1994; Bertin \& Lin 1996). Therefore
later--type spirals have more gas with which to form stars and it therefore
follows that late--type spiral galaxies might have larger specific
star--formation rates than
early--type spiral galaxies. Such a correlation has been shown and discussed
by James et al. (2004), who show that a correlation exists between 
H$\alpha$ equivalent width (which is related to the specific star formation 
rate and the birthrate parameter) and galaxy type. 
Is it therefore possible that the correlation
shown in Fig.\ 1 is affected by a selection bias such as the one described
here? In an attempt to answer this we have investigated the relationship
between shear rate and morphological type (Fig.\ 2), and $L_{FIR}/L_K$ and 
morphological type (Fig.\ 3). Fig.\ 2 shows a weak and insignificant
correlation (correlation coefficient=0.24; significance=70.90 per cent). 
Although it is insignificant the correlation is in the expected sense, 
with early type galaxies have higher rates of shear. The weakness of the
correlation may be attributed to the problems associated with assigning
galaxies with a Hubble type, which is usually a process that is done by
eye and sometimes bears little resemblance to the underlying stellar
mass distribution (Seigar \& James 1998a, b; Seigar et al. 2005).
Fig.\ 3 also shows a weak and insignificant correlation (correlation
coefficient=0.20; significance=83.29 per cent), although once again the
correlation is in the expected sense. Given the weakness, and the low
significance of the correlations shown in Fig.\ 2 and Fig.\ 3, it is
unlikely that the good correlation between shear rate and $L_{FIR}/L_{K}$
is a selection affect.

We believe that the correlation shown in 
Fig.\ 1 is a useful diagnostic tool. Given a
rotation curve and far-infrared data for a galaxy, it should be possible
to measure the stellar mass in any given galaxy, and this could be a very
powerful tool. We now investigate the affect of galaxy size on the overall
star formation rate.

Before going any further, 
one factor should be taken into account when interpreting the results of
this analysis. The psf of IRAS was not capable of resolving most nearby
external galaxies, and this is certainly the case for this sample. In such
an analysis, we are really only interested in the star formation in the
disks of spiral galaxies. However, the use of IRAS fluxes means that the
star formation rates presented in this letter may be contaminated by
bulge and/or nuclear star formation. Furthermore, we have used the total 2MASS
K magnitude, which will also have a significant bulge contribution. One
may expect these two affects to cancel to some degree, although one should
also consider that the bulge contribution to the K-band light is probably
more significant than the bulge contribution to the far-infrared light.
However, since the correlation in
Fig.\ 1 is good, one can only assume that this affect is small
for the current sample. It should be noted that any object with known 
nuclear activity was excluded from the sample.

Using equation \ref{kenn1} we have calculated the star formation rates in 
these galaxies. These are listed in Table 1. 
Since we have measurements of $D_{25}$ for all of these galaxies, which have
been taken from de Vaucouleurs et al. (1991), it is possible to measure
the star formation rate per unit surface area, or the far-infrared luminosity
per unit area, if we assume perfectly 
circular disks. One might expect that
larger galaxies will also be more massive, at least in terms of their stellar
mass. As a result, one might expect to see a correlation between shear 
rate and the far-infrared luminosity per unit area. Fig.\ 4 shows a plot
of shear rate versus the far-infrared luminosity per unit area. Only a very
weak, insignificant correlation is shown (correlation coefficient=0.13; 
significance=47.37 per cent). As a result, we decided to investigate
the relationship between size, parameterized as $D_{25}$ 
and K-band luminosity (or stellar mass) in disk galaxies, and this is shown
in Fig.\ 5. There seems to be
a weak, yet significant, correlation between these two parameters 
(correlation coefficient=0.35; significance=99.50 per cent). As a 
result it seems that while $K$ band luminosity is a good means for estimating
stellar mass (e.g. Brinchmann \& Ellis 2000), it is not so good for the 
overall size of galaxies. Larger galaxies are not necessarily brighter in
the near-infrared. However, on the average, galaxy size and mass do correlate,
but for specific cases, it may not be possible to estimate either mass from
size or size from mass. Also, the lack of a correlation in Fig.\ 4 
may be a result of the weak correlation shown in Fig.\ 5.

From Fig.\ 1, it can be seen that there is a shear rate at which the
star formation in disk galaxies switching off. In fact, from equation 4,
this {\em critical} shear rate is 
$\left(\frac{A}{\omega}\right)_{c}=0.70\pm0.09$. 
Using the correlation between shear rate and spiral arm pitch angle
(Seigar et al. 2005), it can be shown that this critical shear rate
would result in spiral structure with a pitch angle, 
$P_{K}=10^{\circ}\hspace*{-1.3mm}.6\pm2^{\circ}\hspace*{-1.3mm}.5$,
which is consistent with the tightest spiral structure typically seen
in the near-infrared (e.g. Block et al. 1999). This
also suggests that there would be a central mass concentration for which
spiral galaxy disks are stabilised against gravitational collapse
and subsequent star formation.
One would expect that this critical shear rate corresponds to
the regime, where the Toomre stability parameter, $Q=1$. This result 
implies that the correlation shown in Fig.\ 1 does contain some star 
formation physics, as it seems to accurately predict the tightest
wound spiral arms observed in disk galaxies.



\section*{Acknowledgments}

This research has made use of the NASA/IPAC Extragalactic Database (NED) 
which is operated by the Jet Propulsion Laboratory, California Institute of 
Technology, under contract with the National Aeronautics and Space 
Administration. 
This publication makes use of data products from the Two Micron All Sky 
Survey, which is a joint project of the University of Massachusetts and 
the Infrared Processing and Analysis Center/California Institute of 
Technology, funded by the National Aeronautics and Space Administration 
and the National Science Foundation.
The author wishes to thank the referee for useful suggestions
which improved the content of this article.

\end{document}